\documentclass[12pt,a4paper]{article}

\usepackage[includeheadfoot,
            marginratio={1:1,2:3}, 
            width=450pt, 
            height=688pt,]{geometry}

\usepackage{verbatim}
\usepackage{amsmath}
\usepackage{amsfonts}
\usepackage{amssymb}
\usepackage{mathtools}
\usepackage{physics}
\usepackage{siunitx}
\usepackage{graphicx}
\usepackage{tikz} 
\usetikzlibrary{decorations.pathreplacing,calc,calligraphy}
\usepackage{booktabs}
\usepackage{adjustbox}
\usepackage[utf8]{inputenc}
\usepackage{multirow}
\usepackage[splitrule]{footmisc}

\usepackage{empheq}
\usepackage{paralist}
\usepackage{cite}
\usepackage[hidelinks]{hyperref}
\usepackage{xcolor}
\usepackage{tensor}

\usepackage{nicefrac}

\newcommand{\nc}{\newcommand}
\nc{\lb}{\llbracket}
\nc{\rb}{\rrbracket}
\nc{\gl}{\llbracket}
\nc{\gr}{\rrbracket}
\nc{\del}{\partial}
\nc{\tri}{\hspace{-3.5pt}\vartriangle\hspace{-3.5pt}}
\nc{\blacktri}{\blacktriangle}

\nc{\eq}[1]{\begin{equation}
                     \begin{split} #1 \end{split}
                     \end{equation}}
\nc{\ul}{\underline}
\nc{\ov}{\overline}

\nc{\fa}{\hat}
\nc{\fb}{\MakeUppercase}
\nc{\fc}{\tilde}
\nc{\Lie}{{\cal L}} 
\nc{\lambdabar}{{\mkern0.75mu\mathchar '26\mkern -9.75mu\lambda}}

\allowdisplaybreaks[2]
\numberwithin{equation}{section}
\newcommand{\Ls}{\ensuremath{\Lambda_\text{s}}}

\newcommand{\mKK}{\ensuremath{m_\text{KK}}}
\DeclareUnicodeCharacter{2212}{-}
\begin{document}

\pagenumbering{gobble}

\vspace*{-1.5cm}
\begin{flushright}
  {\small
  LMU-ASC 38/23\\
  MPP-2023-281\\
  }
\end{flushright}

\vspace{0.5cm}
\begin{center}
  {\Large \bf
    Starobinsky Inflation in the Swampland 
} 
\vspace{0.2cm}

\end{center}

\vspace{0.15cm}
\begin{center}
Dieter L\"ust$^{1,2}$,
Joaquin Masias$^{2}$,
Benjamin Muntz$^{2,3,4}$,
Marco Scalisi$^{2}$ \\[0.2cm]
\end{center}

\vspace{0.0cm}
\begin{center} 
{\footnotesize
\emph{
$^{1}$ Arnold Sommerfeld Center for Theoretical Physics, Ludwig-Maximilians-Universit\"at M\"unchen, 80333 M\"unchen, Germany
 } 
\\[0.1cm] 
\vspace{0.25cm} 
\emph{$^{2}$  Max-Planck-Institut f\"ur Physik (Werner-Heisenberg-Institut), \\ 
Boltzmannstr. 8, 85748 Garching, Germany}\\[0.1cm]
\vspace{0.25cm} 
\emph{$^{3}$  School of Physics and Astronomy, University of Nottingham, Nottingham, NG7 2RD, UK}\\[0.1cm]
\vspace{0.25cm} 
\emph{$^{4}$  Nottingham Centre of Gravity, Nottingham, NG7 2RD, UK}\\[0.1cm]
}
\end{center} 

\vspace{0.3cm}


\begin{abstract}
\noindent We argue that the Starobinsky model of inflation, realised via an $R^2$ term in the Lagrangian, can originate from quantum effects due to a tower of light species. By means of two separate arguments, we show how this implies that the scale of the $R^2$ term must be of order of the species scale $\Ls$, namely the energy at which gravity becomes strongly coupled. We discuss the implications and challenges of this scenario for inflation, inflationary reheating, and string theory embeddings. In this context, we collect strong evidence to conclude that Starobinsky inflation lies in the Swampland.

\end{abstract}

\clearpage

\pagenumbering{arabic}


\tableofcontents

\section{Introduction}
\label{sec:intro}

Since the 1980s, a plethora of inflation scenarios have been suggested and scrutinised, whilst observational constraints have become increasingly sharpened. Out of these, the Starobinsky model \cite{Starobinsky:1980te} has so far stood the test of time in providing predictions in excellent agreement with observational data \cite{Planck:2018vyg}. Its premise is the addition of a squared curvature term to the standard Einstein-Hilbert action,
\begin{equation}\label{eq:starobinsky}
    S = \int \dd^4 x\ \sqrt{-g}\left[ \frac{M_P^2}{2}\left(R + \frac{R^2}{M^2}\right) \right].
\end{equation}
The scale that controls this correction can be fixed using measurements of the amplitude of scalar perturbations in the Cosmic Microwave Background (CMB), yielding \mbox{$M\simeq \SI{e14}{\giga\electronvolt}$}. In fact, it can be shown that the theory described by eq.~\eqref{eq:starobinsky} is equivalent to a standard Einstein-Hilbert action plus a scalar field with a definite scalar potential, which makes the extraction of phenomenological predictions convenient. Despite the simplicity of this model, it has so far remained a challenge to find a concrete UV embedding (see e.g. \cite{Cecotti:1987sa,Simon:1991bm,Farakos:2013cqa,Ketov:2013sfa,Myrzakulov:2014hca,Broy:2015zba,Asaka:2015vza,Farakos:2015ksa,Cribiori:2023gcy,Brinkmann:2023eph}). Taming curvature contributions of order higher than two has especially proven to be difficult. These terms usually spoil the plateau of the inflaton potential for the case of Starobinsky, thereby also undoing the possibility of having slow-roll for a sufficient number of e-folds (see \cite{Ferrara:2013kca} for a different conclusion in an effective supergravity model).

In this article, we would like to address the following question: is Starobinsky inflation, viewed as an effective field theory (EFT), even consistent with quantum gravity to begin with? Our results point out serious challenges for such compatibility. We do not consider any specific microscopic embedding, but rather employ a general framework and tools proper to the approach of the Swampland programme \cite{Ooguri:2006in,Palti:2019pca,vanBeest:2021lhn}. 

Our strategy and outline are structured in the following way: first of all, we provide compelling evidence in sec.~\ref{sec:R2} that the quadratic term of eq.~\eqref{eq:starobinsky} can originate from the quantum effects of a tower of light species.\footnote{Note that the original paper by Starobinsky \cite{Starobinsky:1980te} contained more second order curvature terms, which were argued to stem from quantum loop corrections of `matter fields' to the Einstein field equations. Therefore, the scale $M$ was expected to depend on the number of these fields (see also \cite{PhysRevD.32.2511} for a more pedagogical account). However, in some subsequent articles \cite{Starobinsky:1983zz,Kofman:1985aw}, Starobinsky proposed a simplified model just with the $R^2$ term, remaining however agnostic about its origin. In sec.~\ref{sec:R2} of this paper, we provide supporting arguments to relate the scale $M$ to the renormalisation effects of towers of species.} A large $N$ number of gravitationally coupled species does indeed contribute to a modification of the scale at which gravity becomes strongly coupled. This is the so-called {\it species scale} \cite{Dvali:2007hz,Dvali:2007wp,Dvali:2009ks,Dvali:2010vm,Dvali:2012uq}, which in $d$ dimensions is given by
\begin{equation}\label{eq:speciesscale}
\Lambda_{\rm s}=\frac{M_P}{N^{\frac{1}{d-2}}}\,.
\end{equation}
It can be interpreted as the renormalization of the Planck mass and thus the scale at which quantum gravitational effects start becoming relevant. The effective decrease of the quantum gravity cutoff, in the limit of large $N$, makes one of the most famous slogans of the Swampland programme very explicit. Indeed, it implies that quantum gravity can be relevant even for EFTs at low energies (namely at energies lower than the Planck scale). In the context of a quasi-de Sitter cosmological phase, it was for example shown that the sole decrease of $\Lambda_{\rm s}$ implies strict limitations on the inflaton range \cite{Scalisi:2018eaz,Scalisi:2019gfv} (see also \cite{vandeHeisteeg:2023uxj} for a more recent work) and also on the gravitino mass \cite{Cribiori:2021gbf}. Other previous analyses of the role of species in the context of inflation were performed in \cite{Antoniadis:2014xva}. String theory provides very concrete examples of scenarios involving a large $N$ number of species and thus a lowered quantum gravity cutoff. Prominent instances of these include the decompactification and/or tensionless string limits \cite{Lee:2018urn,Lee:2019wij}.

Provided that the $R^2$ term in eq.~\eqref{eq:starobinsky} stems from renormalization effects of a tower of light species, we then demonstrate in sec.~\ref{sec:R2} that one can identify the scale $M$ with the species scale up to an order one factor, i.e. $M\simeq \Lambda_{\rm s}$. We deliver two concrete arguments to support this result. One is based on the cosmology of a quasi de-Sitter phase. A second more rigorous explanation is instead based on explicit calculations of the graviton propagator.

With the take-home message of sec.~\ref{sec:R2}, we highlight three immediate implications. One is that, in the Starobinsky model, the scale at which inflation happens is necessarily of the order of the species scale $H\simeq\Ls$. This sets this scenario at the boundary of validity of gravitational weakly coupled EFTs. Second is that \mbox{$M\simeq \SI{e14}{\giga\electronvolt}$} implies an excessive number of species $N\simeq 10^{10}$ present in the early universe, all with masses below the cutoff, which may signal inconsistency of the original 4d effective description. Third, we highlight that generally the species scale is scalar field dependent, and therefore we should not expect $M$ to remain fixed throughout the course of inflation.  In sec.~\ref{sec:inflation} we explore the cosmological consequences of these for CMB inflationary observables. We demonstrate that, for this scenario to align with the current observational constraints on the spectral tilt of scalar perturbations, the species scale $\Ls$ must exhibit a negligible exponential dependency on the inflaton field. Specifically, our analysis reveals that
\begin{equation}
    -0.004\leq \frac{\Ls'}{\Ls}\leq0.001 \,,
\end{equation}
with $\Ls'$ being the derivative along the inflationary trajectory. This result is in clear tension with Swampland bounds presented in the most recent literature, which in general $d$ dimensions, quotes \mbox{$\abs{\Ls'/\Ls}\geq\left[(d-1)(d-2)\right]^{-1/2}$} \cite{Calderon-Infante:2023ler,vandeHeisteeg:2023uxj}. In sec.~\ref{sec:reheating}, we discuss implications for inflationary reheating. We observe that in the Starobinsky scenario, the scale $M$ governs also the physics of reheating happening during the oscillation phase around the minimum of the scalar potential. We show that the identification $M\simeq\Ls$ leads to a strengthening of the upper bound on the reheating temperature $T_{\rm rh}$ by one to two orders of magnitude (depending on the nature of the species). In sec.~\ref{sec:strings} we comment on the species scale and origin of the $R^2$ term in settings of string compactifications.

\section{\texorpdfstring{$R^2$}{R2} gravity and the species scale}\label{sec:R2}

In this section, we would like to provide two convincing arguments suggesting that
\begin{equation}\label{Mspeciesscale}
    M \simeq \Ls\,.
\end{equation}
Put in words, the scale $M$ appearing in the $R^2$ term must be of the order of the scale at which gravity becomes strongly coupled. To preface, let us consider the modified $f(R)$ theory of gravity with action
\begin{equation}\label{eq:R2first}
         S = \dfrac{M_P^2}{2}\int d^4x \sqrt{-g^{(J)}}~ f\left(R^{(J)}\right)=\dfrac{M_P^2}{2}\int d^4x \sqrt{-g^{(J)}} \left(R^{(J)}+\dfrac{(R^{(J)})^2}{M^2}\right)\,,
\end{equation} 
where the superscript is used to distinguish the Jordan frame.\footnote{The reason why this can be referred to as the `Jordan frame' \cite{Mishra:2019ymr,Brinkmann:2023eph} is that the function $f\left(R^{(J)}\right)$ can be rewritten as $\Omega^2(\phi)R^{(J)}-U(\phi)$, with $U(\phi) =M^{-2}(R^{(J)})^2=\frac{1}{4}M^2(1-e^{\sqrt{2/3}\,\phi/M_P})^2$ once we employ eq.~\eqref{eq:Omega}. Namely, this gives a Lagrangian linear in $R^{(J)}$ multiplied by a function of a field $\phi$ with a scalar potential $U(\phi)$. Here we have adopted the definition of Jordan frame as being such that the metric couples minimally to matter fields $\psi$ appearing in the Lagrangian $\mathcal{L}_m(g_{\mu\nu},\psi)$, which for the sake of simplicity we omit in our investigation.} To write the action in the Einstein frame, one can perform a conformal transformation 
\begin{equation}
    g_{\mu\nu}^{(J)} \to g_{\mu\nu}^{(E)} = \Omega^2g_{\mu\nu}^{(J)}
\end{equation}
and promote the conformal factor to a dynamical scalar field that is gravitationally coupled (it will then also couple to the matter sector). Identifying 
\begin{equation}\label{eq:Omega}
    \Omega^2 =\frac{\partial f\left(R^{(J)}\right)}{\partial R^{(J)}} = 1 + 2\frac{R^{(J)}}{M^2}= \exp\left(\sqrt{\frac{2}{3}}\frac{\phi}{M_P}\right)
\end{equation}
brings the action into Einstein frame in terms of the metric $g_{\mu\nu}^{(E)}$ and the canonically normalised field $\phi$. That is
\begin{equation}\label{eq:staropot}
     S = \int d^4x \sqrt{-g^{(E)}} \left[\dfrac{M_P^2}{2} R^{(E)} +\dfrac{1}{2}g_{\mu\nu}^{(E)}\partial^\mu \phi\partial^\nu \phi - \dfrac{M^2M_P^2}{8}\left(1-e^{-\sqrt{\frac{2}{3}}\phi/M_P}\right)^2 \right]\,,
\end{equation}
where one can recognise the typical Starobinsky plateau potential with the overall amplitude fixed by the scale $M$, which gives predictions in excellent agreement with observations.

In this section, we would like to provide two different arguments that the scale $M$ is naturally generated by the quantum effects of a tower of species. This will allow us to make the identification eq.~\eqref{Mspeciesscale}.

\subsection{Cosmological argument}
During inflation one has $e^{-\sqrt{\frac{2}{3}}\,\phi/M_P}\ll 1$, such that the potential term in eq.~\eqref{eq:staropot} behaves as a cosmological constant. In this regime the theory is thus approximately equivalent to that of gravity with a cosmological constant and a single modulus. Interpreting it this way, the cosmological constant is related to the scale of the $R^2$ correction as
\begin{equation}\label{eq:cc}
    \Lambda=\dfrac{M^2M_P^2}{8}\,.
\end{equation}
Furthermore, the scalar field is related to the curvature by
\begin{equation}\label{eq:fieldcuvaturerelation}
    \sqrt{\frac{2}{3}}\frac{\phi}{M_P} = \ln\Omega^2 = \ln\left(1+2\frac{R^{(J)}}{M^2}\right)\,.
\end{equation}
The slow-roll regime is therefore that of high Jordan frame curvature $R^{(J)}\gg M^2$ and $\Omega^2\gg 1$. We require instead the Einstein frame curvature to be upper bounded by the species scale, $R^{(E)}\lesssim \Ls^2$, as we have argued in the introduction that $\Ls$ can be considered a cutoff for standard Einstein gravity. From the conformal transformation, we have the relation
\begin{equation}\label{eq:Rscalarsrelation}
    R^{(J)} = \Omega^2\left( R^{(E)} + \frac{1}{M_P^2}g_{\mu\nu}^{(E)} \partial^\mu \phi \partial^\nu \phi + \frac{1}{M_P}\sqrt{6}\Box^{(E)}\phi\right) \approx \Omega^2R^{(E)}\,,
\end{equation}
where the approximation is valid when the potential energy of the field dominates. From eq.~\eqref{eq:fieldcuvaturerelation}, we can infer that the conformal factor in the slow-roll limit is related to the Einstein frame curvature by
\begin{equation}\label{eq:omegaRrelation}
    \Omega^2\simeq \left(1-2\dfrac{R^{(E)}}{M^2}\right)^{-1}\,.
\end{equation}
Therefore, for an Einstein frame curvature saturating the upper bound $R^{(E)}\simeq\Ls^2$, demanding that $\Omega^2$ remains positive requires
\begin{equation}
   1-2\dfrac{\Ls^2}{M^2}>0\,.
\end{equation}
This leads to a lower bound on $M$,
\begin{equation} 
\label{eq:Mlower}
    M^2\gtrsim 2\Ls^2\,.
\end{equation}
During inflation, on the other hand, the Starobinsky model describes a theory of Einstein-Hilbert gravity with a cosmological constant \eqref{eq:cc} and curvature
\begin{equation}
    R^{(E)}\simeq\dfrac{4\Lambda}{ M_P^2}=\dfrac{M^2}{2}\,.
\end{equation}
This result can be found by tracing the Einstein field equation with negligible stress-energy contribution. Again, requiring $R^{(E)}\lesssim \Ls^2$ yields
\begin{equation}
    M^2\lesssim 2\Ls^2\,.
     \label{eq:Mupper}
\end{equation}
Combining equations \eqref{eq:Mlower} and \eqref{eq:Mupper} therefore suggests that
\begin{equation}
    M^2\simeq 2\Ls^2\,. 
\end{equation}

\subsection{Perturbative argument}
In this subsection, we demonstrate that the correction to the graviton propagator given by a tower of $N$ light species is identical to that of an $R^2$ term in the Lagrangian.

\subsubsection{Graviton propagator for \texorpdfstring{$R$}{R} + tower of species}
A spectrum of $N$ light species coupled to gravity produces a quantum one-loop correction to the graviton propagator
\begin{equation}\label{eq:gravitonpropagator}
    \Pi^{\mu\nu\rho\sigma}(p^2) = K^{\mu\nu\rho\sigma}\pi(p^2)\,,\qquad K^{\mu\nu\rho\sigma}= \dfrac{1}{2}\left( \eta^{\mu\rho}\eta^{\nu\sigma} + \eta^{\mu\sigma}\eta^{\nu\rho} - \eta^{\mu\nu}\eta^{\rho\sigma} \right)\,.
\end{equation}
Crucially, we refer to the tensorial and momentum independent part $K^{\mu\nu\rho\sigma}$ as the \emph{index structure} of the propagator. It satisfies the property
\begin{equation}
  K^{\mu\nu\rho\sigma}  h_{\rho\sigma}=h^{\mu\nu}-\dfrac{1}{2}\eta^{\mu\nu} h\,,
\end{equation}
where the right hand side is the trace-reversed variable commonly used in linearized gravity. The momentum dependent piece is given by
\begin{equation}\label{eq:propagatormomentum}
    \pi^{-1}(p^2) =  p^2\left[1- \frac{N p^2}{120\pi M_P^2}\log\left( -\frac{p^2}{\mu^2} \right) \right]\,,
\end{equation}
where $\mu$ is the renormalization scale. The energy scale at which the EFT breaks down due to gravity becoming strongly coupled is precisely when the one-loop correction becomes comparable with the tree-level term. This is what we expect to be the species scale. Solving for the cutoff momentum squared in terms of the Lambert W function gives \cite{Castellano_2023, Calmet_2016}
\begin{equation}
    p^2 = 120\pi\frac{M_P^2}{N}\left[ W_{-1}\left( -\frac{120\pi M_P^2}{\mu^2 N} \right) \right]^{-1}\overset{!}{\simeq} \Ls^2\,.
\end{equation}
In principle the definition of the species scale only holds up to logarithmic corrections. In the large $N$ limit, $W_{-1}(-1/N)=\log(1/N)+\log\log\text{corrections}$, so up to numerical factors,
\begin{equation}\label{eq:speciesNlogN}
\begin{split}
    \Ls &\sim \frac{M_P}{\sqrt{N}} \left[ W_{-1}\left( -\frac{M_P^2}{\mu^2N} \right) \right]^{-\frac{1}{2}}\\
    &\sim \frac{M_P}{\sqrt{N\log N}}\,.
\end{split}
\end{equation}
As such, the logarithmic correction to the species scale has an inherently quantum origin in that they stem from the $\log(-p^2/\mu^2)$ divergence due to the one-loop renormalization. We will comment more on this at the end of the section. But for the sake of brevity, treating the logarithmic divergence as correction to the definition of the species scale, we may therefore write the one-loop corrected graviton propagator as
\begin{equation}\label{eq:propsimple}
    \pi^{-1}(p^2) \simeq  p^2\left(1- \dfrac{p^2}{\Ls^2} \right)\,.
\end{equation}
In the following, we argue that $\Ls$ is coincident to the scale $M$ controlling the $R^2$ term in the Starobinsky model. Our argument is based on the fact that the graviton propagator has the same index structure and diverges at exactly the same energy scale, when one identifies $M\simeq \Ls$.

\subsubsection{Graviton propagator for \texorpdfstring{$R+R^2$}{R+R2}}
For now, let us drop superscripts and work entirely in terms of the Jordan frame. Consider then the variation of the action \eqref{eq:starobinsky} with respect to the metric \cite{Arapoglu:2010rz,Cooney:2009rr}
\begin{equation}
\begin{split}
         \delta S = \dfrac{M_P^2}{2}\int \dd^4x \sqrt{-g}\  &\bigg(R_{\mu\nu}-\dfrac{1}{2}g_{\mu\nu}R\\
         &+\dfrac{1}{M^2}\Big[2R_{\mu\nu} R-g_{\mu\nu} R^2-2(\nabla_{\mu}\nabla_{\nu}-g_{\mu\nu}\Box)R\Big]\bigg)\delta g^{\mu\nu}\,.
\end{split}
\end{equation}
The vanishing of the bracket gives the Einstein field equations. Next, we consider metric perturbations around a flat background\footnote{One could carry out the same background metric analysis and arrive at the same conclusion by expanding around de Sitter, employing the results in \cite{Alvarez-Gaume:2015rwa,Hell:2023mph}.}
\begin{equation}
    g_{\mu\nu}=\eta_{\mu\nu}+h_{\mu\nu}\,.
\end{equation}
Up to leading order in $h_{\mu\nu}$, the linearized field equations read
\begin{equation}
    \left(\dfrac{1}{2}K_{\mu\nu\rho\sigma} - \frac{1}{M^2}\left( \partial_\mu\partial_\nu - \Box\eta_{\mu\nu}\right)\eta_{\rho\sigma}\right)\Box h^{\rho\sigma}=0\,,
\end{equation}
where $K_{\mu\nu\rho\sigma}$ is given in eq.~\eqref{eq:gravitonpropagator}. This equation of motion, which has a different index structure than pure Einstein-Hilbert without the $R^2$ correction, can also be obtained from the variation of the action
\begin{equation}\label{eq:starobinskylagrangian}
S= \frac{M_P^2}{4} \int d^4x \sqrt{-g}\ h^{\mu\nu}    \left(\dfrac{1}{2}K_{\mu\nu\rho\sigma} - \frac{1}{M^2}\left( \partial_\mu\partial_\nu - \Box\eta_{\mu\nu}\right)\eta_{\rho\sigma}\right)\Box h^{\rho\sigma}.
\end{equation}
An important thing to note at this point is that the propagator in eq. \eqref{eq:gravitonpropagator} is written using the harmonic gauge \cite{Gasperini:2017ggf},
\begin{equation}
    \partial_\mu h^{\mu}_\nu=\dfrac{1}{2}\partial_\nu h\,,\qquad \partial_\mu\partial_\nu h^{\mu\nu}=\dfrac{1}{2}\Box h\,.
\end{equation}
This allows us to rewrite all derivatives in the Lagrangian in terms of d'Alembert operators
\begin{equation}\label{eq:starobinskylagrangian2}
S= \frac{M_P^2}{8}\int d^4x \sqrt{-g}\ h^{\mu\nu}    \left(K_{\mu\nu\rho\sigma} + \frac{\Box}{M^2} \eta_{\mu\nu}\eta_{\rho\sigma}\right)\Box h^{\rho\sigma}.
\end{equation}
To appropriately interpret the propagator of this theory, we need to decompose eq. ($\ref{eq:starobinskylagrangian2}$) in terms of graviton projectors
\begin{equation}
    S=  \dfrac{M_P^2}{8}\int d^4x \sqrt{-g} h^{\mu\nu}\left[P^{(2)}_{\mu\nu\rho\sigma}\Box  -\dfrac{P^{(0)}_{\mu\nu\rho\sigma}}{2}\left(\Box-6\dfrac{\Box^2}{M^2}\right) \right]h^{\rho\sigma}\,,
\end{equation}
where
\begin{equation}
\begin{split}
    P^{(2)}_{\mu\nu\rho\sigma}&=\dfrac{1}{2}\left(\eta_{\mu\rho}\eta_{\nu\sigma}+\eta_{\nu\rho}\eta_{\mu\sigma}\right)-\dfrac{1}{3}\eta_{\mu\nu}\eta_{\rho\sigma}\,,\\
    P^{(0)}_{\mu\nu\rho\sigma}&=\dfrac{1}{3}\eta_{\mu\nu}\eta_{\rho\sigma}
\end{split}
\end{equation}
are the so-called Barnes-Rivers projectors. More specifically, $P^{(2)}_{\mu\nu\rho\sigma}$ projects onto the spin-2 transverse traceless part of $h^{\rho\sigma}$, whilst $P^{(0)}_{\mu\nu\rho\sigma}$ projects onto the spin-0 component. We can now directly obtain the propagators for the spin-2 and spin-0 modes by going to momentum space and inverting the momentum dependent parts
\begin{equation}
\begin{split}
    \Pi^{(2)}_{\mu\nu\rho\sigma}(p^2)&=\dfrac{P^{(2)}_{\mu\nu\rho\sigma}}{p^2}\,,\\
    \Pi^{(0)}_{\mu\nu\rho\sigma}(p^2)&=\dfrac{1}{2}P^{(0)}_{\mu\nu\rho\sigma}\left(p^2-\dfrac{6p^4}{M^2}\right)^{-1}=\dfrac{P^{(0)}_{\mu\nu\rho\sigma}}{2}\left[\left(p^2-\dfrac{M^2}{6}\right)^{-1}-\dfrac{1}{p^2}\right]\,.
\end{split}
\end{equation}
 Adding both propagators we find
\begin{equation}
       \Pi^{(2)}_{\mu\nu\rho\sigma}(p^2)+   \Pi^{(0)}_{\mu\nu\rho\sigma}(p^2)= K_{\mu\nu\rho\sigma}\frac{1}{p^2}+\dfrac{1}{6}\eta_{\mu\nu}\eta_{\rho\sigma}\left(p^2-\dfrac{M^2}{6}\right)^{-1}\,,
\end{equation}
where the two terms correspond respectively to the vacuum graviton propagator and a scalar degree of freedom with mass $m_\phi = M/\sqrt{6}$. The latter can also be observed by expanding the Einstein frame action \eqref{eq:staropot} around the minimum of the potential at $\phi = 0$,
\begin{equation}
     S = \int d^4x \sqrt{-g} \left(\dfrac{M_P^2}{2}R +\dfrac{1}{2}g^{\mu\nu}\partial_\mu \phi\partial_\nu \phi - \dfrac{1}{2}\dfrac{M^2}{6}\phi^2 + \cdots \right)\,.
\end{equation}

\subsubsection{Graviton propagator for \texorpdfstring{$R+\mathcal{O}(\mathcal{R}^2)$}{R+OR2}  -- fixing the index structure}
In order to relate the scale $M$ to the species scale $\Ls$, we need to also recover the same index structure of the propagator as in eq.~\eqref{eq:gravitonpropagator}. We shall show that by introducing other quadratic curvature terms in the action 
\begin{equation*}
    R^2\,,\qquad R_{\mu\nu}R^{\mu\nu}\,,\qquad R_{\mu\nu\rho\sigma}R^{\mu\nu\rho\sigma}\,,
\end{equation*}
i.e. all curvature operators of mass dimension 4, we may modify the graviton propagator in order to get the overall factor $K^{\mu\nu\rho\sigma}$. This rightfully so seems to steer away from the point of analysing the Starobinsky model. But in four dimensions and in an FLRW background, as a good approximation in the slow-roll regime, these additional corrections can be grouped into topological terms.

Let us start by providing, in harmonic gauge, the second order contributions coming from linearized squared curvature terms.
\begin{equation}
\label{eq:R2termssplitlinear}
\begin{split}
    R^2 &= \dfrac{1}{4}h\Box^2 h\,,\\
     R_{\mu\nu}R^{\mu\nu} &= \dfrac{1}{4}h^{\mu\nu}\Box^2 h_{\mu\nu}\,,\\
    R_{\mu\nu\rho\sigma}R^{\mu\nu\rho\sigma} &= h^{\mu\nu}\Box^2 h_{\mu\nu} -\dfrac{1}{4}h\Box^2 h\,. 
\end{split}
\end{equation} 
The Einstein-Hilbert term gives a contribution of the form \eqref{eq:starobinskylagrangian} but with the standard index structure $K_{\mu\nu\rho\sigma}$. Hence, to second order in metric perturbations, a general gravitational action including squared curvature corrections can be expressed as
\begin{subequations} \label{eq:ActionR2general}
\begin{align}
    S &= \dfrac{M_P^2}{2}\int \dd^4 x\ \sqrt{-g} \left[R+\dfrac{1}{M^2}\left(R^2+c_2R_{\mu\nu}R^{\mu\nu}+c_3 R_{\mu\nu\rho\sigma}R^{\mu\nu\rho\sigma}\right)\right]\,, \label{eq:squaredaction}\\
\begin{split} 
    &= \dfrac{M_P^2}{8} \int \dd^4x\ h^{\mu\nu}\bigg[K_{\mu\nu\rho\sigma}\\
    &\qquad - \dfrac{2}{M^2}\left((-\frac{c_2}{2}-2c_3)\frac{1}{2}\left(\eta_{\mu\rho}\eta_{\nu\sigma} + \eta_{\nu\rho}\eta_{\mu\sigma}\right) - (1-c_3)\frac{1}{2}\eta_{\mu\nu}\eta_{\rho\sigma}\right)\Box \bigg]\Box h^{\rho\sigma}\,.
\end{split}
\end{align}
\end{subequations}
In order to obtain a correction to the propagator with the same index structure we require that
\begin{equation}\label{eq:coefficients}
    1-c_3=-\frac{c_2}{2}-2c_3=\alpha\implies \begin{cases}
        c_2 = 2\alpha-4\\
        c_3 = 1-\alpha
    \end{cases}
\end{equation}
for some constant $\alpha$. Demanding that the coefficients satisfy eq.~\eqref{eq:coefficients} for general values of $\alpha$, the linearized action becomes
\begin{equation}
     S = \dfrac{M_P^2}{8}\int \dd^4x\ h^{\mu\nu}K_{\mu\nu\rho\sigma}\left(\Box-\dfrac{2\alpha}{M^2}\Box^2\right)h^{\rho\sigma}\,.
\end{equation}
We can now easily read the correction to the propagator
\begin{equation}\label{eq:propagatorR2species}
    \pi^{-1}(p^2)\simeq p^2\left(1-\dfrac{2\alpha}{M^2}p^2\right).
\end{equation}
Hence, it is evident that, in a theory with quadratic curvature corrections, gravity becomes strongly coupled at  the scale
\begin{equation}
p^2=\frac{M^2}{2\alpha}\,.
\end{equation}
The propagator of this theory has the same index structure and pole of the one-loop corrected propagator derived in the case of a tower of species, see eq.~\eqref{eq:propsimple}. If one requires that the pole of both propagators is at the same energy scale, one finds 
\begin{equation}
    M\simeq\Ls\,.
\end{equation}

\noindent Two final comments are in order. Firstly, it is well-known that $f(R)$ theories with ${\partial_{R^2} f(R)\neq 0}$ generally suffer from unitarity violation due to the presesence of propagating ghosts \cite{Briscese:2013lna, PhysRevD.16.953}. When allowing the gravitational action to include general squared curvature terms and arbitrary $\alpha$, like eq.~\eqref{eq:ActionR2general}, we may potentially need to worry about this. However, for the specific case of $d=4$ dimensions, the Gauss-Bonnet term is topological and appears only up to a surface term. One can therefore always remove $R_{\mu\nu\rho\sigma}R^{\mu\nu\rho\sigma}$ from the action by expressing it in terms of $R^2$ and $R_{\mu\nu}R^{\mu\nu}$. In fact, the case $\alpha=0$ is exactly when eq.~\eqref{eq:ActionR2general} gives Gauss-Bonnet. Similarly, in an expanding cosmology such as an FLRW spacetime, it is known that the combination 
\begin{equation}
     \int \dd^4x\ \sqrt{-g} \left(R^2-3R_{\mu\nu}R^{\mu\nu}\right)
\end{equation} 
is also topological \cite{Briscese:2012ys,Briscese:2013lna}. Thus, one can express the general squared curvature action entirely in terms of $R^2$. An alternative way to see this is by expressing the general action \eqref{eq:squaredaction} in a terms of $R^2$, Gauss-Bonnet, and the Weyl tensor squared $C_{\mu\nu\rho\sigma}C^{\mu\nu\rho\sigma}$. Since the Weyl tensor vanishes on conformally flat backgrounds, it should not lead to the propagation of ghosts during inflation. At much lower energies, when the Einstein-Hilbert term dominates over the $\mathcal{O}(\mathcal{R}^2)$ terms, ghosts also should not pose a problem as long as the suppression scale accompanying the squared curvature terms is sufficiently large, as it is the case when one identifies ${M\simeq \Ls\simeq \Lambda_\text{UV}}$ (a similar line of though has been presented in \cite{Brax:2023nvx}).

It may then seem counter-intuitive to claim that we can recover the index structure  \eqref{eq:gravitonpropagator} from eq.~\eqref{eq:ActionR2general} if $R^2$ is the only non-trivial contribution to the gravitational EFT. This can be made clear by looking at the FLRW metric in conformal time \mbox{$g_{\mu\nu}=a(t)\eta_{\mu\nu}$} with $a(0)=1$. Metric perturbations around a flat background at time $t=\epsilon$ are given by
\begin{equation}
    h_{\mu\nu}=(a(\epsilon)-1)\eta_{\mu\nu}\,,\qquad h=\eta^{\mu\nu}h_{\mu\nu}=4(a(\epsilon)-1)\,,
\end{equation}
such that the linearized metric can be expressed solely in terms of its trace through \mbox{$h_{\mu\nu}=\frac{1}{4}h\,\eta_{\mu\nu}$}. The kinetic energy term associated to  eq. (\ref{eq:gravitonpropagator}) takes the form 
\begin{equation}
 h_{\mu \nu}   K^{\mu\nu\rho\sigma}\pi(p^2)h_{\rho\sigma}=-\dfrac{1}{4}\pi(p^2)h^2,
\end{equation}
such that the effective index structure is given by
\begin{equation}
    \widetilde{K}^{\mu\nu\rho\sigma}= -\dfrac{1}{4}\eta^{\mu\nu}\eta^{\rho\sigma}\,.
\end{equation}
As can be read from eq. (\ref{eq:R2termssplitlinear}), this is precisely the effective index structure associated to the $R^2$ term. We thus see that, in an FLRW universe, the index structure associated to the Einstein-Hilbert term together with the tower of states, matches that of the $R^2$ term.\\[12pt]
Finally, it is worth returning our attention to the logarithmic divergence discussed at the beginning of this section.  Whilst we have presented this as a multiplicative correction in this simplified estimate \eqref{eq:speciesNlogN}, the precise algebraic arrangement is still a topic of ongoing discussion. In particular, eq.~\eqref{eq:propagatormomentum} is computed in a massless approximation. But it has also been argued in \cite{Cribiori:2023ffn,Cribiori:2023sch,vandeHeisteeg:2022btw,Cribiori:2022nke,Basile:2023blg} that a more careful computation of the species scale using massive fields yields instead \emph{additive} logarithmic corrections. Furthermore, it was highlighted in \cite{Cribiori:2023sch} that multiplicative logarithmic corrections break modular invariance. Even disregarding these points, one can acknowledge that logarithmic terms such as this may generally be regarded as quantum in origin -- in particular when involving a scale of renormalization. Our goal was to show that a curvature term like $R^2/M^2$ gives rise to the same correction to the propagator in eq.~\eqref{eq:propagatormomentum}, and we have done this at the level of linearized gravity. Hence, we of course cannot expect to capture this logarithmic feature using what is essentially a background field approach. It is instead recovered when we recall that the Starobinsky model in our discussion should be strictly treated as an EFT. In that case $M$ is to be regarded as an effective coupling that runs with the energy scale \cite{Donoghue:1994dn, Donoghue:2022eay, Calmet:2016fsr, Copeland:2013vva}. The necessary factor of $\log(-p^2/\mu^2)$ is thus provided by the graviton loop renormalization.\footnote{Note that it is only loops of massive particles and not gravitons that contribute to the renormalization of $M_P$. Therefore the Einstein-Hilbert term is not affected.}
In a Barvinsky and Vilkovisky fashion \cite{Barvinsky:1985an, Barvinsky:1990up, Barvinsky:1993en, Barvinsky:1995jv}, one can choose to equivalently represent this effect by adding a non-local term to the effective action 
\begin{equation}\label{eq:R2nonlocal}
    S = \frac{M_P^2}{2}\int\dd^4 x\ \sqrt{-g}\left( R+R\left[ \frac{1}{M^2} + b\log\left( \frac{\Box}{\mu^2}\right)\right]R \right).
\end{equation}
It is easy to see how this gives the factor $\log(-p^2/\mu^2)$, discussed earlier in this section, and has the correct index structure, once we go to momentum space. Here the value of $b$, related to the beta function (see e.g. \cite{Calmet:2016fsr} for an explicit computation), is fully determined by the IR structure of gravity \cite{Espriu:2005qn}. $M$ thus depends on $\mu$ in such a way that the total action is independent of $\mu$. But what is most important to notice is that this presentation further supports the argued claims that logarithmic corrections should be additive and not multiplicative. The addition of non-local terms of course modify the Starobinsky model and has cosmological consequences. These have been studied for instance in \cite{Espriu:2005qn, Donoghue:2014yha, Cabrer:2007xm, Calmet:2016fsr}. In this context, let us also mention a related study \cite{Edholm:2016seu} which considers an action similar to eq.~\eqref{eq:R2nonlocal} containing an infinite series of non-local squared curvature terms, organised in powers of $\Box/M_{\rm nl}^2$. From cosmological constraints it is then argued that the scale of non-locality is bounded by $M_{\rm nl}\gtrsim\mathcal{O}(10^{14})\SI{}{\giga\electronvolt}$. We find it interesting that this lower bound happens to align with what we argue to be the quantum gravitational cutoff of the theory, namely the species scale $\Ls\simeq M\simeq \SI{e14}{\giga\electronvolt}$.

\section{Consequences for inflation}\label{sec:inflation}

In the previous section, we have provided two independent arguments to identify the scale of the $R^2$ corrections with the species scale, i.e. $M\simeq\Ls$. There are at least three immediate implications of this identification:
\begin{enumerate}
    \item By looking at eq.~\eqref{eq:staropot} or eq.~\eqref{eq:cc}, we observe that the scale of $R^2$ corrections also determines the scale of inflation $H$. Indeed, we find that $M\simeq H$. Together with the result of sec.~\ref{sec:R2}, this implies that
    \begin{equation}
    H\simeq\Ls \,,
    \end{equation}
    suggesting that the Starobinsky scenario, as a model of inflation, is {\it at best at the boundary of validity} of a weakly coupled gravitational EFT.
  
    \item Given that normalization of the CMB scalar power spectrum requires $M\simeq \SI{e14}{\giga\electronvolt}$ \cite{Planck:2018vyg}, one can use eq.~\eqref{eq:speciesscale} to infer the corresponding number of gravitationally coupled species with mass below the cutoff present in the early universe. In $d=4$, this turns out to be $N\simeq 10^{10}$. If this was to come for instance from a single tower of Kaluza-Klein (KK) modes, the associated KK mass scale would have to be
    \begin{equation}
    m_{\rm KK}\simeq 10^4~{\rm GeV}\,.
    \end{equation}
    The fact that KK modes appear already at ten orders of magnitude below the scale of inflation suggests that Starobinsky cannot be regarded as a consistent 4d EFT. 

    \item Over super-Planckian field excursions, we generally expect that the species scale drops exponentially with respect to the inflaton field. If so, this implies a parametric dependency \begin{equation}\label{eq:Mexp}
    M \sim e^{-\gamma\phi/M_P},
    \end{equation}
    where $\gamma$ is an order one number. The above follows from the Swampland Distance Conjecture (SDC) \cite{Ooguri:2006in}, which dictates that the mass scale of a tower of species decreases exponentially towards asymptotic regions in field space. This behaviour is expected to be observed at least after a field variation of order $M_P$ \cite{Baume:2016psm,Klaewer:2016kiy}. Since towers of species determine the species scale, one can demonstrate that the SDC implies also an exponential drop-off of $\Ls$ \cite{Hebecker:2018vxz,Scalisi:2018eaz,Castellano_2023} and therefore eq.~\eqref{eq:Mexp}. This immediately imposes an effective modification to the standard Starobinsky inflaton potential written in eq.~\eqref{eq:staropot}.
\end{enumerate} 
In the following, we will focus on the last point and investigate the consequences of the exponential decay of the scale $M$, given in  eq.~\eqref{eq:Mexp}, on the inflationary dynamics and observables. We will consider first a scenario in which the exponential behaviour of $M$ spans for the entire period of inflation. We will then consider a scenario where $M$ is given as a piecewise function in order to model the effective take-over of the exponential behaviour at one point during inflation.

\subsection*{Scenario 1}

One possible situation is that the scale $M$ exhibits the exponential functional behaviour throughout the whole period of inflation. The scalar potential eq.~\eqref{eq:staropot} will be then effectively modified as
\begin{equation}\label{eq:potgamma}
   V(\phi)=\dfrac{M_P^2M_*^2}{8}e^{2\gamma\phi/M_P}\left(1-e^{-\sqrt{\frac{2}{3}}\phi/M_P}\right)^2\,,
\end{equation}
with $M_*$ being a constant. The sign of $\gamma$ has been set such that positive values correspond to an exponential drop-off of the scale $M$ towards the end of inflation.

Provided the scalar potential above, we can now compute the scalar spectral tilt and tensor to scalar ratio up to leading corrections in $\gamma$. One can demonstrate that the magnitude of $\gamma=\abs{\Ls'/\Ls}$ in this class of models (i.e. scalar potentials composed as $V(\phi)=e^{2\gamma\phi/M_{P,d}} f(\phi)$ with $f(\phi)$ being a function allowing for successful slow-roll inflation) must be upper bounded by a quantity smaller than unity in order to deliver cosmic acceleration (see a demonstration at this end of this section). This argument justifies why we will be able to expand the inflationary observables in a limit where $\gamma$ is small. 

\begin{figure}[t!]
    \centering
        \begin{tabular}{c c}
        \includegraphics[width=0.5\textwidth]{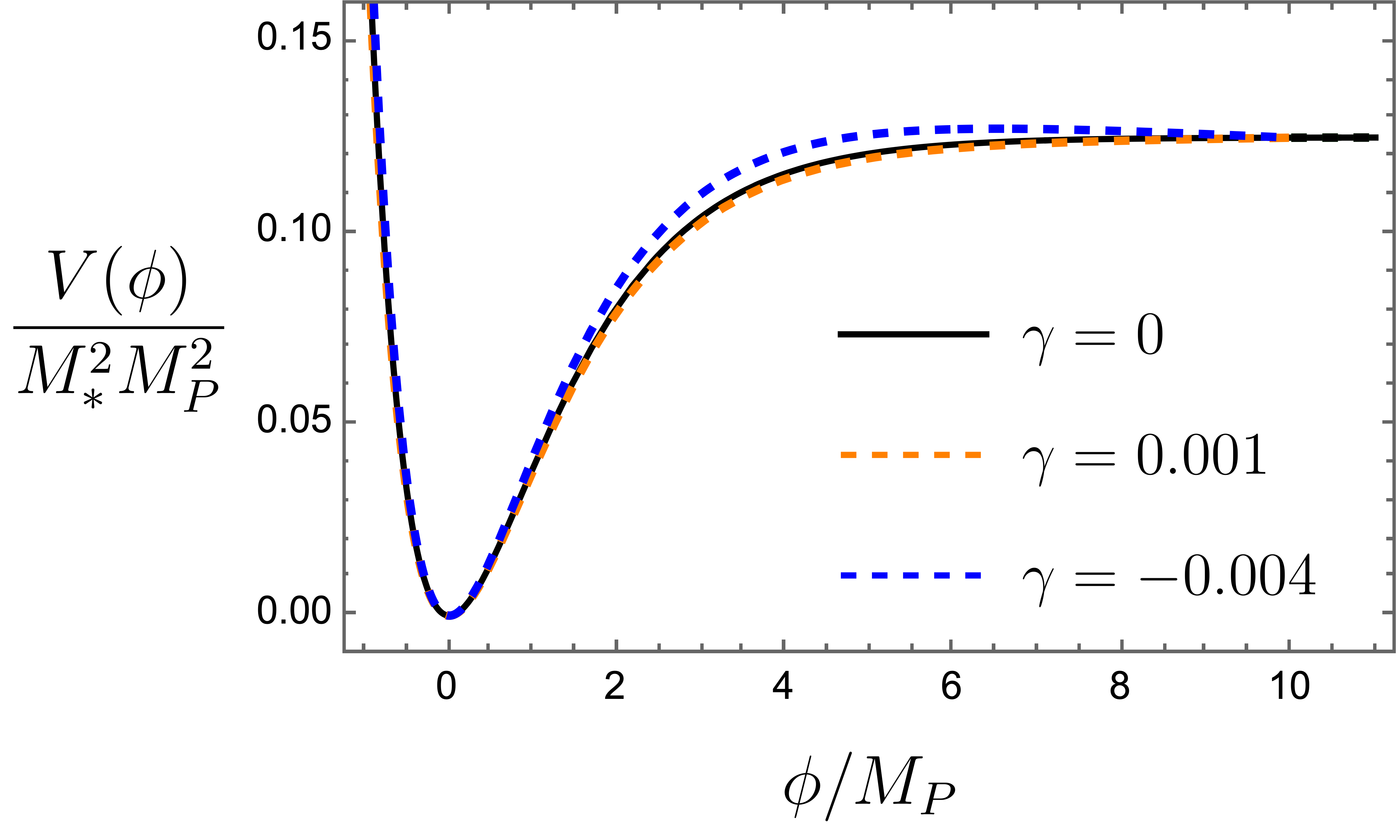} & \includegraphics[width=0.45\textwidth]{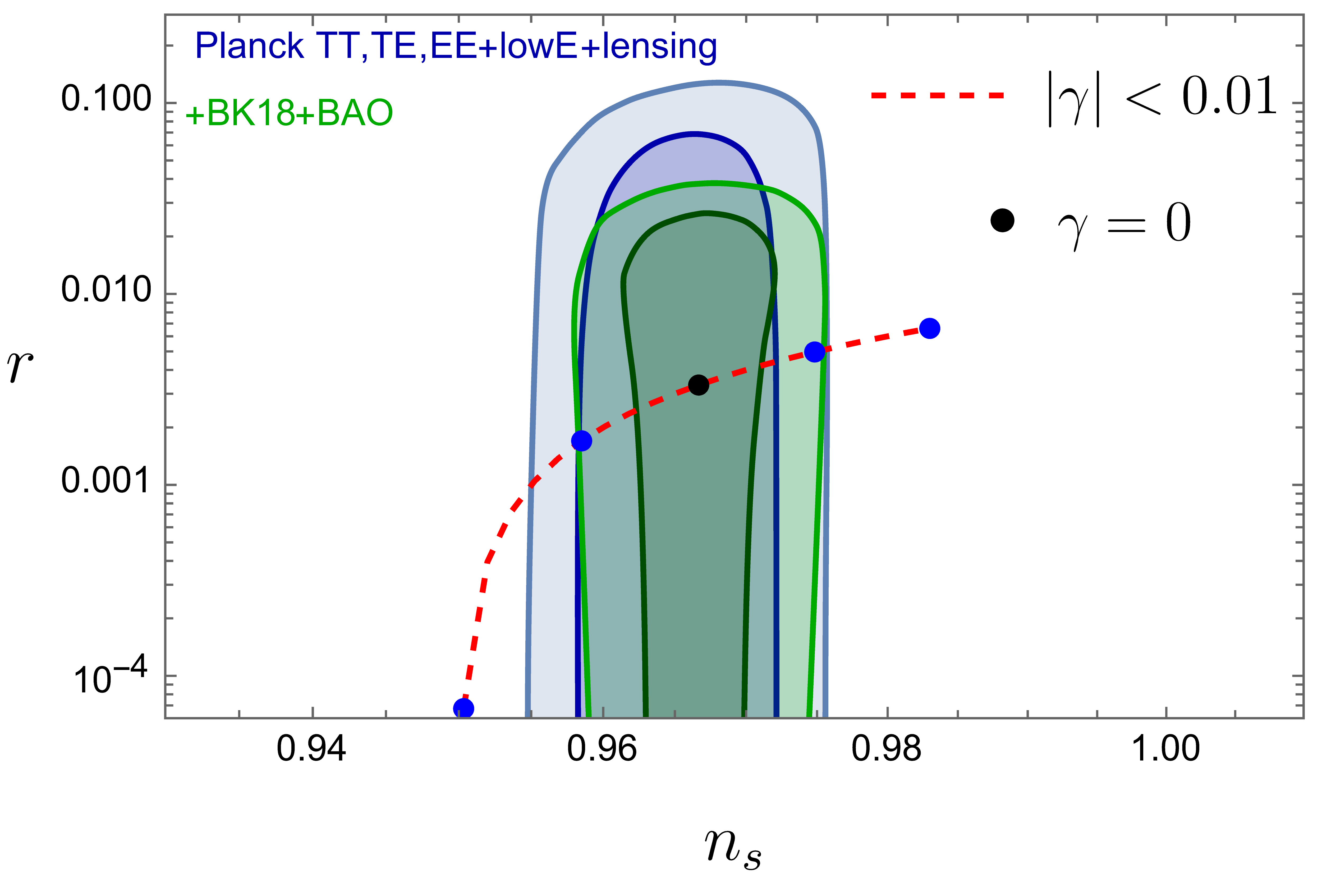} 
    \end{tabular}
    \caption{Left: {\it Starobinsky potential compared to corrections due to an exponential dependence of the scale $M$.} Right: {\it Cosmological observables $n_s$ vs. $r$ with  ${\abs{\gamma}<0.01}$ in models described by eq.~\eqref{eq:potgamma}. The points in blue denote benchmark values ${\gamma=\{-0.01,-0.005,0.005,0.01\}}$.}}
    \label{fig:staroplot}
\end{figure}

One can then calculate the scalar power spectrum and the slow-roll parameters of this scenario (see Appendix \ref{sec:appendixinflationobservables} for the calculation details) to obtain the corrections to the scalar spectral tilt and the tensor-to-scalar ratio. They yield
\begin{equation}
\begin{split}
    n_s-1&\simeq -\frac{2}{N_e} + 2 \gamma \sqrt{\frac{2}{3}} + \mathcal{O}(\gamma^2)\,,\\[10pt]
    r&\simeq \frac{12}{N_e^2}+\gamma \dfrac{8 \sqrt{6}}{N_e} + \mathcal{O}(\gamma^2)\,,
    \label{eq:nsr1}
\end{split}
\end{equation}
with $N_e$ being the total number of e-folds during inflation. Observational constraints \cite{Planck:2018vyg} for the value of the spectral tilt require $n_s-1= -0.035\pm 0.004$, which translates to a range for the exponential decay rate
\begin{equation}
-0.004\leq\gamma\leq0.001\,.
\label{eq:gammarang}
\end{equation}
As depicted in fig.~\ref{fig:staroplot}, potentials that reproduce the bounds on $n_s$ cannot deviate significantly from the Starobinsky potential (black line with $\gamma=0$). The above range for $\gamma$ violates a lower bound recently pointed out in literature \cite{Calderon-Infante:2023ler,vandeHeisteeg:2023uxj}
\begin{equation} \label{eq:gammalower}
    \abs{\gamma}\geq\dfrac{1}{\sqrt{(d-2)(d-1)}}=\dfrac{1}{\sqrt{6}}\,,
\end{equation}
with the last equality holding in $d=4$. This result provides further indication that Starobinsky inflation should be in the Swampland.

\subsection*{Scenario 2}

In known EFTs of string theory, the species scale exhibits a manifest exponential behaviour near the boundary of the moduli space. There one generally expects infinite towers of species becoming light and falling below the cutoff. In the interior of the moduli space, $\Ls$ usually exhibits instead a different functional behaviour with smaller slope. It also approaches a maximum in one point in field space, which has been refereed to as `centre of the moduli space' and thus corresponds to a minimum number of light species. See \cite{vandeHeisteeg:2022btw,Cribiori:2022nke,vandeHeisteeg:2023ubh,Andriot:2023isc,Cribiori:2023sch,vandeHeisteeg:2023dlw} for some recent works about this topic.

In this scenario, we consider that there exists a point $\phi_*$ in field space where the transition to the exponential behaviour of $M$ takes over. We treat the scale $M=M(\phi)$ as a piecewise function so to schematically have
\begin{equation}
    M(\phi) = \begin{cases}
    M_*\,, & \phi \geq \phi_\star \vspace{0.4cm}\\
    M_* e^{-\gamma(\phi_\star-\phi)/M_P}\,, & \phi < \phi_\star
    \end{cases}\,,
\end{equation}
where the constant $M_*$ mimics the behaviour of the species scale in the interior of the moduli space. The scalar potential eq.~\eqref{eq:staropot} will be modified accordingly,
\begin{equation}
    V(\phi) = \begin{cases}
    \dfrac{M_P^2M_*^2}{8}\left(1-e^{-\sqrt{\frac{2}{3}}\phi/M_P}\right)^2\,, & \phi \geq \phi_\star \vspace{0.6cm}\\
    \dfrac{M_P^2M_*^2}{8}e^{2\gamma(\phi-\phi_\star)/M_P}\left(1-e^{-\sqrt{\frac{2}{3}}\phi/M_P}\right)^2\,, & \phi < \phi_\star
    \end{cases}\,.
\end{equation}
In this case, inflation spans two subsequent phases: first it undergoes slow-roll as prescribed by the standard Starobinsky potential, followed by the modified potential, which lasts all the way until the end of inflation. This could be interpreted as the inflation starting somewhere in the bulk interior of moduli space (where we approximate the species scale to be a constant) and ending towards the boundary (where the exponential behaviour takes over). The point $\phi_*$ can likewise be considered as a free parameter describing `how much' inflation takes place in the asymptotic region of moduli space. Also in this case, the corrections to the spectral tilt and tensor to scalar ratio can be calculated and turn out to be
\begin{equation}
\begin{split}
    n_s-1&= -\frac{2}{N_e}+\gamma\dfrac{ 3 \sqrt{\frac{3}{2}} e^{\sqrt{\frac{2}{3}} \phi_*/M_P } \left(e^{\sqrt{\frac{2}{3}} \phi_*/M_P }-1\right)}{4 N_e^2} + \mathcal{O}(\gamma^2)\,,\\[12pt]
    r&=\frac{12}{N_e^2}+\gamma\frac{9 \sqrt{\frac{3}{2}}  e^{\sqrt{\frac{2}{3}} \phi_*/M_P } \left(e^{\sqrt{\frac{2}{3}} \phi_*/M_P }-1\right)}{N_e^3} + \mathcal{O}(\gamma^2)\,.
\end{split}
\end{equation}
Here we have assumed the point in field space where inflation ends $\phi_\text{end}\simeq 0$ such that the limit $\phi_*\to 0$ recovers the expected Starobinsky behaviour. In the limit $\phi_*\to\phi_{N_e}$, where
\begin{equation}
    \phi_{N_e}\simeq \sqrt{\frac{3}{2}} \log\left(\frac{4N_e}{3}\right)M_P\simeq 5.4~ M_P
\end{equation}
is the value in field space at which there is $N_e$ e-folds before the end of inflation (in the last equality we have assumed $N_e=60$), we instead recover the predictions eq.~\eqref{eq:nsr1}. 

\begin{figure}[t!]
    \centering
        \begin{tabular}{c c}
        \includegraphics[width=0.5\textwidth]{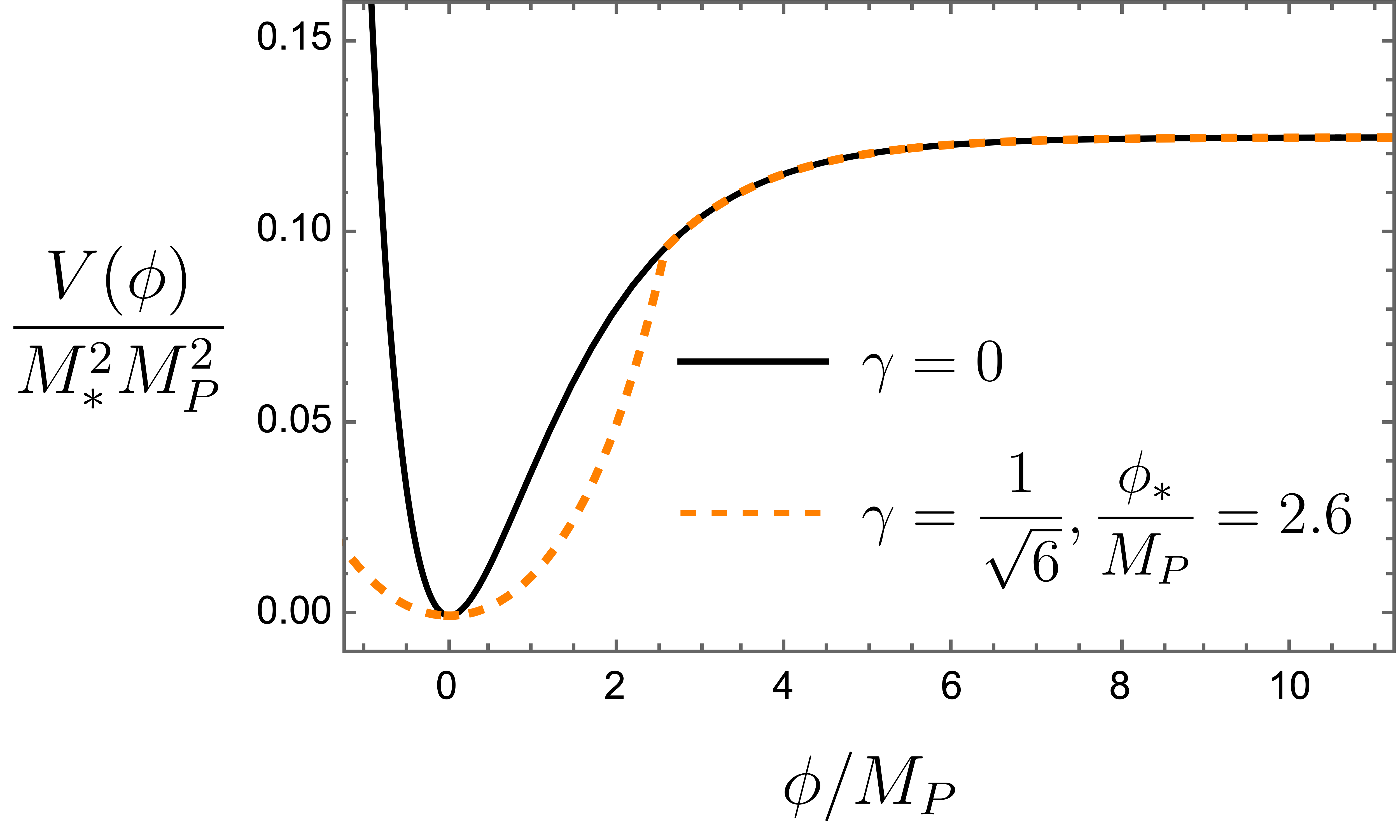} & \includegraphics[width=0.45\textwidth]{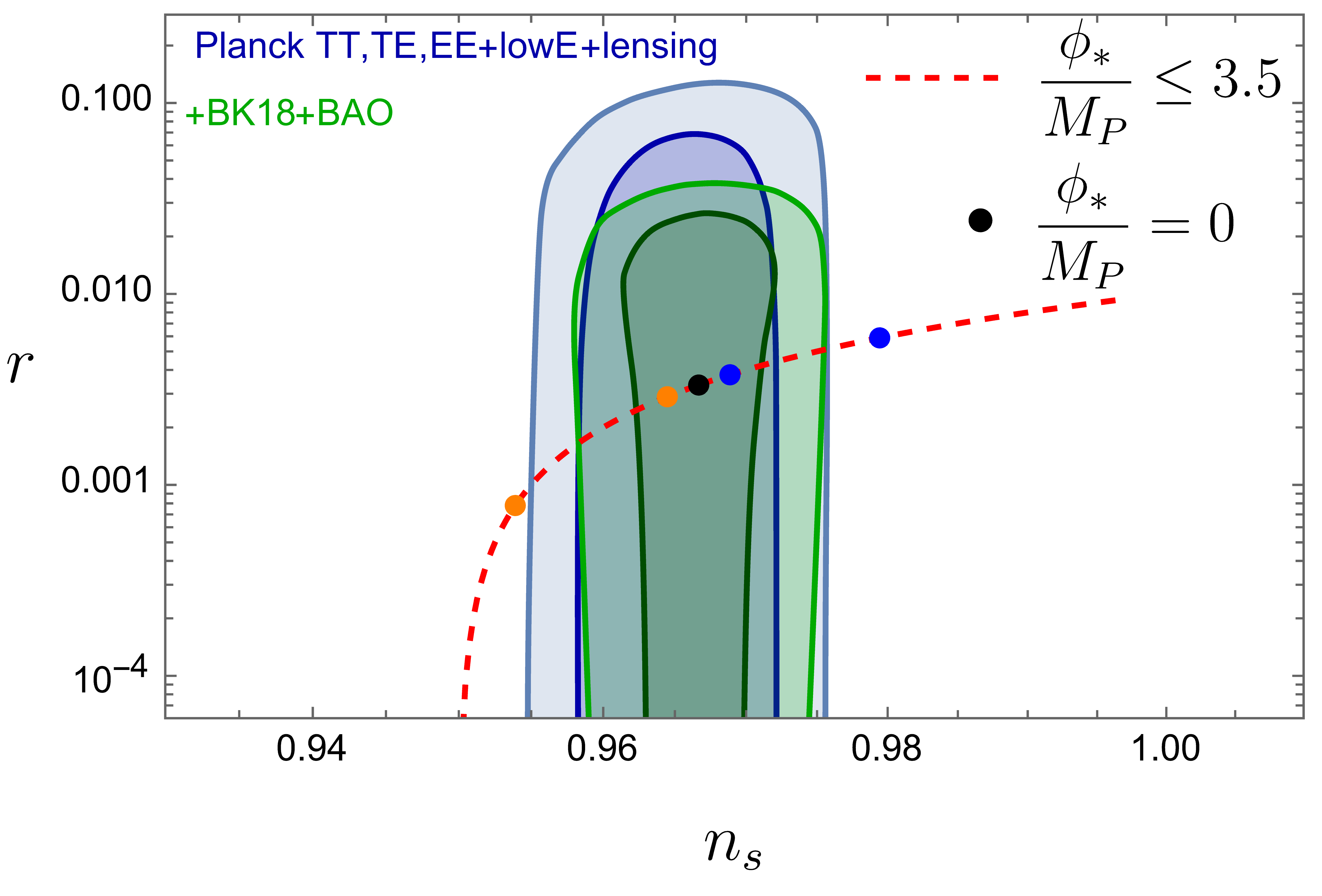} 
    \end{tabular}
    \caption{Left: {\it Starobinsky potential compared to possible corrections due to an exponential dependence of the scale $M$ with $\gamma=\frac{1}{\sqrt{6}}$  and $\phi_*<\phi_{N_e}$.} Right: {\it Cosmological observables $n_s$ vs. $r$ for  and $\phi_*\leq 3.5$, and exponential scaling $\gamma=\pm\frac{1}{\sqrt{6}}$, such that positive (negative) values of $\gamma$ correspond to blue (orange) points. These denote benchmark values $\phi_*/M_P=\{2,3\}$.}}
    \label{fig:staroplot2}
\end{figure}

Within this scenario, it is possible to increase the decay rate $\gamma$ such to make it compatible with the theoretical constraints as given in eq.~\eqref{eq:gammalower}, namely $|\gamma|\geq 1/\sqrt{6}$ in four dimensions. To obtain compatibility with CMB observations one needs to consider 
\begin{equation}
   0.38\lesssim \dfrac{\phi_*}{\phi_{N_e}}\lesssim0.47\,.
\end{equation}
Hence, for $N_e=60$, one has that the exponential behaviour of $M$ starts, at the earliest, at
\begin{equation}
    \phi_*\simeq 2.6~ M_P\,
    \label{eq:distmin}
\end{equation}
corresponding to an initial field displacement 
\begin{equation}
\phi_{N_e}-\phi_*\simeq 2.8~ M_P\,.
    \label{eq:distmin}
\end{equation}
during which the scale $M$ (and hence the species scale) remains constant. This potential is depicted in fig.~\ref{fig:staroplot2}, where the start of the exponential behaviour acts as a stark drop-off (this is more evident due to the increased value of $\gamma$). Such a modest super-Planckian field displacement is potentially in tension with the SDC, which suggests that before this can happen, $\Ls$ should start to exponentially decay.

\subsection*{Upper bound on \texorpdfstring{$\gamma=\abs{\Ls'/\Ls}$}{Ls'/Ls}}

To obtain the range of values of $\gamma$ required for slow-roll, let us consider a $d$-dimensional theory of scalar field inflation
\begin{equation}
         S = \int \dd^d x \sqrt{-g}\left[ \dfrac{M_{P,d}^{d-2}}{2} R+\frac{1}{2}g^{\mu\nu}\partial_\mu\phi\partial_\nu\phi-V(\phi)\right]\,.
\end{equation}
Note that $M_{P,d}$ is the $d$-dimensional Planck mass. The Friedmann equations and the equation of motion for the inflaton take the form
\begin{align}
    H^2 &=\dfrac{2}{(d-1)(d-2)}\left(\dfrac{1}{2}\dot\phi^2+V(\phi)\right) M_{P,d}^{-(d-2)}\,,\\
    \dot{H} &=\dfrac{1}{d-2}\dot\phi^2 M_{P,d}^{-(d-2)}\,,\\
    0&=\ddot{\phi}+(d-1) H\dot{\phi}+V'(\phi)\,.
\end{align}
Using these results we obtain a standard expression for the slow-roll parameter in terms of the scalar potential,
\begin{equation}
    \varepsilon\equiv-\dfrac{\dot{H}}{H^2}\approx \dfrac{d-2}{4}\left(\dfrac{V'(\phi)}{V(\phi)}\right)^2 M_{P,d}^2\,.
\end{equation}
Consider now the scenario where inflation occurs due to the scalar field rolling towards some smaller values, as is the case for Starobinsky inflation with our convention. Suppose the scalar potential takes the general form
\begin{equation}\label{eq:Vexpmodification}
    V(\phi)=e^{2\gamma\phi/M_{P,d}} f(\phi)\,,
\end{equation}
such that $f$ admits a flat region $\abs{f'/f}\ll 1$ to allow for slow-roll phase. The slow-roll parameter for the potential $V(\phi)$ reads
\begin{equation}
    \varepsilon= \dfrac{d-2}{4}\left(2\gamma+\dfrac{f'}{f}\right)^2 \simeq (d-2)\gamma^2\,.
\end{equation}
A phase of cosmic acceleration requires $\varepsilon<1$. This translates immediately into an upper bound on the slope of the scale $M$, hence of the species scale, such as
\begin{equation}\label{eq:gammabound}
    \abs{\gamma}= \abs{\frac{\Ls'}{\Ls}}<\dfrac{1}{\sqrt{d-2}}\,.
\end{equation}
Bigger values of $\gamma$ would spoil the flatness of the potential. Interestingly, this bound has also been pointed out in some more stringy settings in \cite{vandeHeisteeg:2023ubh, vandeHeisteeg:2023dlw} (see also \cite{Castellano:2023stg,Castellano:2023jjt, Rudelius:2023spc}).

\section{Consequences for reheating}\label{sec:reheating}

In the previous section, we have already pointed out that the identification $M\simeq\Ls$ together with the value $M\simeq \SI{e14}{\giga\electronvolt}$, imposed by CMB observations, implies the existence of a very large number of species $N\sim 10^{10}$. A concrete possibility, motivated by string theory, is that these would be KK modes. In this section we will explore whether such a numerous number of massive modes spell doom for particle processes in the early universe.

After inflation, the energy associated to the inflaton is transferred to other relevant light degrees of freedom in the universe. This leads to an epoch of radiation domination and eventually to the production of atomic nuclei, also known as Big Bang nucleosynthesis (BBN). This process in which the universe is in approximate thermal equilibrium, is commonly referred to as reheating. Since the temperature at the beginning of reheating can be comparable to the scale of inflation, a UV cutoff $\Ls$ close to that same scale can have strong phenomenological implications. 

First we should note, despite the fact that even small values of the exponential factor $\gamma$ can have strong effects on the slow-roll in Starobinsky, the value of $M$ is not expected to be significantly modified. For the range found in eq.~\eqref{eq:gammarang}, a field excursion $\Delta\phi\simeq \phi_{N_e}\sim \log(N_e)$ (up to some $\mathcal{O}(1)$ numerical factors), gives an exponential decay $e^{-\gamma\Delta\phi}\simeq 1$. Even in the case $\gamma=1/\sqrt{6}$ ($\gamma$ saturating the theoretical lower bound), eq.~\eqref{eq:distmin} implies that $e^{-\gamma\Delta\phi}\simeq 0.4$. Therefore we generally expect the scale $M$ to be approximately the same between the eras of inflation and reheating.
 
After inflation, the potential oscillates around its minimum at $\phi=0$, and the inflaton acquires a mass
\begin{gather}
m_\phi = \frac{M}{\sqrt{6}}\,.
\end{gather}
Since the thermalization of the universe occurs as a result of the decay of the inflaton, the  temperature at the beginning of reheating can be approximated as
\begin{equation}\label{eq:temp}
    T_\text{rh}\simeq \left(\dfrac{90}{g_* \pi^2}\right)^{1/4}\alpha_\phi\sqrt{m_\phi M_P} \lesssim 10^{-3} M_P\,,
\end{equation}
where $g_*$ is the number of relativistic degrees of freedom and $\alpha_\phi$ is related to the total coupling of the inflaton to light particles
\begin{equation}
    \alpha_\phi\simeq \sum_{i=1}^{g_{*}}\dfrac{g^2_{\phi,i}}{\sqrt{8\pi}}\,,
\end{equation}
with $g_{\phi,i}$ being the coupling constant between the inflaton and each degree of freedom. The upper bound on the reheating temperature of eq.~\eqref{eq:temp} is given by measurements of tensor perturbations \cite{Kofman:1997yn,Lozanov:2019jxc}. From this, one can approximate the Hubble rate during reheating in terms of the reheating temperature as
\begin{equation}
    H\simeq \frac{T_\text{rh}^2}{M_P}=\left(\dfrac{90}{g_* \pi^2}\right)^{1/2}\alpha_\phi^2 m_\phi\ll m_{\phi}\,.
\end{equation}
 Where we consider $g_*\simeq 10^2$ as is the case for the Standard Model, and $g_{\phi,i}^2\ll1$, such that the inflaton is weakly coupled to SM fields. The fields in the tower could in principle contribute to the relativistic degrees of freedom, $g_*$, in the following we make sure this is not the case. We use the simplest model of linearized gravity in order to compute the elastic scattering of some massive scalars $\chi$ via graviton exchange. At some general temperature $T$ and in the relativistic limit, one finds
\begin{equation}
    \sigma\simeq \dfrac{T^2}{M_P^4}\,,
\end{equation}
which holds for particles with mass $m_\chi < T$. This expression is in agreement with similar estimates found in previous literature \cite{Arkani-Hamed:1998sfv, Law-Smith:2023czn, Gonzalo:2022jac}. Heavier modes decouple from the thermal bath, as they cannot be produced and decay to lighter modes. This computation relies on the na\"ive interpretation of $M_P$ as the quantum gravity scale, and is only valid for $T\ll\Ls$. Thus we should in general expect $T_\text{rh}<\Ls\simeq 10^{-5}M_P$. 

Next we consider the interaction rate and freeze-out of these light degrees of freedom. Considering just the interactions between a single mode, this is given up to some numerical factors by 
\begin{equation}
    \Gamma=T^3\sigma\simeq \dfrac{T^5}{M_P^4}\,,
\end{equation}
which is also consistent with the results one could expect from dimensional analysis. Let us define $\hat{N}(T)$ as the number of species with masses below $T$. Considering for instance KK modes in a theory describing $p$ compact and isotropic extra dimensions, this is found to be
\begin{equation}
    \hat{N}(T)=\left(\dfrac{T}{\mKK}\right)^p.
\end{equation}
Each of these states will then in principle have $\hat{N}(T)$ different interaction channels. The total interaction rate gives
\begin{equation}
    \Gamma_\text{total}\simeq \hat{N}(T)\Gamma=\dfrac{T^{5+p}}{\mKK^p M_P^4}=\dfrac{T^{5+p}}{\Ls^{2+p}M_P^2}\,.
\end{equation}
If these KK modes participate in counting the number of relativistic degrees of freedom $g_*$, this could inject an undue contribution to the total coupling $\alpha_\phi$, spoiling reheating predictions. It is therefore important to check whether this is the case. Namely, the tower does not contribute to the light degrees of freedom if their interactions are frozen out. I.e. $\Gamma_\text{total}<H$, which implies
\begin{equation}
     T_\text{rh}<\Ls^{\frac{2+p}{3+p}}M_P^{\frac{1}{3+p}}\,.
\end{equation}
For $\Ls\simeq M$ as argued in this paper, this provides the following constraints on the temperature at the time of reheating, for a single KK $(p=1)$ or string tower ($p=\infty$)\footnote{A string tower presents exponential degeneracy, which is commonly approximated by KK tower with infinite degeneracy.}
\begin{equation}
\begin{split}
    T_\text{rh}&\lesssim 10^{-4}~M_P \qquad (p=1),\\
    T_\text{rh}&\lesssim 10^{-5} M_P \qquad (p=\infty).
    \label{eq:kkmodeprod}
\end{split}
\end{equation}
We found the bounds to be slightly stronger than that imposed by tensor perturbations, eq.~\eqref{eq:temp}. 

Let us likewise comment on whether the presence of a tower of states will spoil BBN. This will crucially depend on the value of $g_*$ at the beginning of radiation domination, where the temperature is $T\simeq \SI{1}{\mega\electronvolt}$. Since $\Gamma_\text{tower}\ll H$ scales with the temperature at a faster rate than $H$, we do not expect the tower to affect BBN.

Most constraints from \cite{Arkani-Hamed:1998sfv} rely on the KK modes being thermalized, while we avoid that by requiring the tower to be decoupled at the time of reheating. We must, however, still ensure that cooling during reheating is mainly due to the expansion of the universe, and not production of KK modes escaping into the additional dimensions. The rate of cooling due to the expansion of the universe is given by
\begin{equation}
    \left(\dfrac{\dd\rho}{\dd t}\right)_\text{expansion}= -3 H\rho\simeq-3\dfrac{T^6}{M_P}\,.
\end{equation}
Similarly, let us estimate the ``evaporation'' rate of energy in the universe. This is mediated by the production of KK modes, which then decay into the additional dimensions, effectively taking energy away from the $d=4$ universe. It is given by \cite{Arkani-Hamed:1998sfv} 
\begin{equation}
    \left(\dfrac{\dd \rho}{\dd t}\right)_\text{evaporation}  \simeq -N(T)\dfrac{T^7}{M_P^2}\simeq\dfrac{T^{7+p}}{m_\text{KK}^{p}M_P^2}\simeq\dfrac{T^{7+p}}{\Ls^{2+p}}\,.
\end{equation}
Having a subdominant evaporation contribution requires
\begin{equation}
    T_\text{rh}<\Ls^{\frac{2+p}{1+p}}M_P^{-\frac{1}{1+p}}.
\end{equation}
This corresponds to
\begin{equation}
\begin{split}
    T_\text{rh}&\lesssim 10^{-8}~M_P \quad (p=1),\\
    T_\text{rh}&\lesssim 10^{-5}~M_P \quad (p=\infty).
\end{split}
\end{equation}
We can conclude that the presence of a tower, through particle production and bulk decay, places stronger bounds on the reheating temperature compared to that of the observational constraint placed in eq. \eqref{eq:temp}.

\section{Comments on species scale, string- and KK towers}\label{sec:strings}
So far we have argued that the $R^2$ term can be interpreted as a quantum gravity correction due to a tower of light species and subsequently presented how this leads to tensions between Starobinsky inflation and Swampland constraints. However, we can still ask how this term may arise in certain UV settings. In this section, we discuss two string theory literature cases in which one can recover the $R^2$ term in 4 dimensions and show that one can consistently identify the scale $M$ with the species scale.
In particular, we consider compactifying the actions of Type IIB and heterotic string theory from $D=10$ dimensions to $d=4$, taking into account leading order $\alpha'$ corrections. Subsequently we attempt to identify the $R^2$ term and scale $M$ in the resulting EFT. Details about these calculations and different conventions are found in Appendix \ref{app:strings}. One finds the following:
\begin{equation}
\begin{split}
    \text{Heterotic:}\qquad M &\simeq M_s\,,\\[10pt]
    \text{Type IIB:}\qquad M &\simeq g_s\frac{M_P}{\mathcal{V}_{s}^{1/6}}\,.
\end{split}
    \label{eq:scaleM}
\end{equation}
Note that $M_P$ is the 4d Planck mass, $g_s$ the string coupling, and $\mathcal{V}_s$ is the dimensionless volume of the entire internal manifold measured in units of the string scale $M_s$.

Next we would like consider what kinds of towers can possibly lead to a species scale that matches these expressions. Adopting the Emergent String Conjecture \cite{Lee:2019wij} for infinite distance limits, the options comprise of either a tower of KK modes of an internal $p$-cycle with volume $\tau_{p;s}$  or massive modes of the fundamental critical string. Up to possible logarithmic corrections, the respective species scales are given by
\begin{equation}
\begin{split}
    \text{KK tower:}\qquad &\left\{\begin{array}{l}
    \Ls \simeq M_{P,10}\simeq \dfrac{M_P}{\sqrt{\tau_{p;s}}}\\[20pt]
    N \simeq \tau_{p;s}
    \end{array}\right.\,,\\[20pt]
    \text{String tower:}\qquad & \left\{\begin{array}{l}
    \Ls \simeq M_s\simeq g_s M_P\\[10pt]
    N\simeq g_s^{-2}
    \end{array}\right.\,.
\end{split}
\end{equation}
In the heterotic case, one can identify the leading tower responsible for the normalization of the scale $M$ with a tower of string states. In this case, the identification $M\simeq\Ls$ which we have argued in this article, follows automatically. For Type IIB, we find two ways of relating $M$ to the species scale, depending on the nature of the tower. Either it can be realised as a string tower for finite volume where $\mathcal{V}_s^{1/6}$ is approximately of order unity. Alternatively, one can identify a tower of KK modes associated to an internal 2-cycle as the leading tower determining the species scale. The 2-cycle volume scales with the overall volume modulus as $\tau_{2;s}\sim \mathcal{V}^{1/3}_s$ thus leading to the identification of the species scale with the scale $M$ in Type IIB as given in eq.~\eqref{eq:scaleM}. The relation between $R^2$ corrections and leading 2-cycle KK tower was also noted in \cite{Cribiori:2022nke}, in the context of higher curvature corrections to black hole entropy. 

Despite all of this, let us emphasise that we do not present a mechanism by which the Starobinsky model is protected from higher curvature corrections. Terms such as $R^n$ for $n>2$ still remain ruinous for the inflationary plateau \cite{Huang:2013hsb,Ivanov:2021chn,Brinkmann:2023eph}, and so far there is no known top-down construction where these can be neglected without fine-tuning. This requires a priori an additional suppression scale parametrically larger than $M$, thus also larger than $\Ls$.

String theory nevertheless provides many more candidates for scalar-tensor theories of inflation. K\"ahler moduli inflation in particular can generate potentials which closely resemble Starobinsky, and can in principle be recast as pure tensor EFTs \cite{Brinkmann:2023eph, Burgess:2016owb}. These models are however unable to reproduce the universal Yukawa coupling $y_\phi=-1/6$ between the inflaton and Standard Model fermions that arises from $f(R)$ theories \cite{Aparicio:2008wh}. As such, a concrete embedding of Starobinsky inflation is still beyond what can be currently achieved from string theory compactifications. Our analysis indeed confirms that purely considering $R+R^2$ is in tension with Swampland conjectures.

\section{Conclusion}

In this article we have argued that the Starobinsky model of inflation, realised via an additional $R^2$ term in the Lagrangian, can be generated by the renormalization effects of a tower of light species. This has led us to conclude that the scale $M$ controlling squared curvature operators, must be of the same order as the species scale $\Ls$, above which gravity becomes strongly coupled. We presented in sec.~\ref{sec:R2} two different and independent arguments to support this claim. In particular, following the perturbative argument, we have shown that, once $M\simeq \Ls$, the one-loop corrected graviton propagator in a theory with $\mathcal{O}(\mathcal{R}^2)$ operators has the same index structure and pole of that obtained in standard Einstein gravity containing a tower of light species \cite{Dvali:2007hz,Dvali:2007wp}.

We have then studied the implications of the identification $M\simeq \Ls$ for cosmology. This in fact leads to a series of important challenges for the model of Starobinsky. First of all, because $M\simeq H$ (see eq.~\eqref{eq:cc}), it implies that the Hubble scale of inflation $H$ must be of the order of the cutoff of the theory. The Starobinsky model as an EFT description is therefore at best at the boundary of validity when describing weakly coupled gravity. Second, given the value $M\simeq \SI{e14}{\giga\electronvolt}$ imposed by observations at CMB scales, it implies a very large number $N\simeq 10^{10}$ of light species. Such a numerous spectrum can lead to severe phenomenological consequences and potentially invalidate the 4d effective theory. Finally, it leads to an effective modification of the original Starobinsky scalar potential, since the scale $M$ (due to its relation to $\Ls$) acquires an exponential dependence on the inflaton field $\phi$ for large scalar field variations, ${M\sim \exp(-\gamma \phi)}$. This is a consequence of the Swampland Distance Conjecture \cite{Ooguri:2006in}. We have explored the consequences of this overall exponential dependence for CMB inflationary observables in sec.~\ref{sec:inflation}. In particular, we have calculated the corrections to scalar spectral index $n_s$ and the tensor to scalar ratio $r$, demonstrating that compatibility with current observational bounds implies $\abs{\gamma} = \abs{\Ls'/\Ls}\ll 1/\sqrt{6}$. Namely, it manifestly violates a theoretical lower bound for the normalized slope of the species scale pointed out in recent literature \cite{Calderon-Infante:2023ler,vandeHeisteeg:2023uxj}. In this context, we have also shown that we are able to recover the theoretical upper bound on $\abs{\gamma}$ appearing in \cite{vandeHeisteeg:2023ubh, vandeHeisteeg:2023dlw}, by simply allowing for successful inflation of this class of models. In general $d$ dimensions we recover $\abs{\gamma}= \abs{\Ls'/\Ls} <(d-2)^{-1/2}$. 

As a further cosmological implication, we have studied inflationary reheating in sec.~\ref{sec:reheating}. In this class of models, the species scale indeed governs the entire cosmological evolution from the beginning of inflation to the oscillatory phase of reheating. The scale $M$ then sets the mass of the quadratic expansion of the scalar potential around the minimum placed at $\phi=0$. We show that the presence of a tower of light species strengthens the upper bound on the reheating temperature  $T_{\rm rh}$ by one to two orders of magnitude, depending on whether KK or string modes compose the tower. 

Finally, in sec.~\ref{sec:strings}, we offered some comments concerning string theory embeddings. We argue that in heterotic string compactifications the leading tower responsible for setting the scale $M$ must be a string tower. Type IIB compactifications additionally allows the leading tower to be identified with KK modes  of a Calabi-Yau 2-cycle. Both cases provide confirmation that the scale of $R^2$ is nothing but the species scale.\\[8pt]
In conclusion, we have provided a number of strong indications supporting the claim that the Starobinsky model of inflation must be in the Swampland.\\[8pt]
It is important to point out that our analysis concerns inflation scenarios whose features stem from higher curvature corrections. Models where a similar (or equal) scalar potential originates from other effects should in principle not be affected by these conclusions. One such example is given by the $\alpha$-attractor models, which feature an extended inflationary plateau with exponential deviation from it -- thus very analogous to that of the Starobinsky model. $\alpha$-attractors \cite{Kallosh:2013yoa} also provide excellent agreement with observational data. However, their origin is related to deformations of the no-scale supergravity structure \cite{Roest:2015qya,Scalisi:2015qga} and in particular to the curvature of the internal K\"ahler manifold. One may wonder if the peculiar properties of the species scale also affect this class of models. We leave this, together with other intriguing questions, for future work. 

\section*{Acknowledgements}

We would like to thank Ed Copeland, Niccol\`o Cribiori, Anamaria Hell, Renata Kallosh, Sergey Ketov, Osmin Lacombe, Andrei Linde, Swagat Mishra, and Tony Padilla for helpful comments and discussions. The work of D.L. is supported by the Origins Excellence Cluster and by the German-Israel-Project (DIP) on Holography and the Swampland. B.M. is supported by a UK Science and Technology Facilities Council studentship.

\section*{Note added}
On the day when this paper appeared on the arXiv, we were informed about the very tragic and unexpected passing of Alexei Starobinsky. We were extremely saddened by his passing and the unfortunately timed appearance of our paper. We wish to emphasise that our work holds no intention of discrediting any of Starobinsky's ideas or contributions. The phrase `in the Swampland' just indicates that the model in question is challenging to reconcile with the swampland conditions on a coherent formulation of quantum gravity. Without any doubt, Alexei Starobinsky is one of the pioneers of cosmic inflation. His seminal contributions to cosmology will last forever.

\appendix
\section{Corrections to primordial inflationary observables}\label{sec:appendixinflationobservables}
The dimensionless scalar power spectrum can be expressed as
\begin{equation}
    P_\zeta=\dfrac{H^2}{8\pi^2 \varepsilon M_P^2}\,,
\end{equation}
where $H$ is the Hubble parameter
\begin{equation}
    H^2=\dfrac{1}{3M_P^2}\left(\dot\phi^2+V(\phi)\right)\simeq \dfrac{V(\phi)}{3M_P^2}\,,
\end{equation}
and $\varepsilon$ is the slow-roll parameter
\begin{equation}
    \varepsilon=-\dfrac{\dot H}{H^2}\simeq\dfrac{1}{2}\left(\dfrac{V'(\phi)}{V(\phi)}\right)^2 M_P^2\,.
\end{equation}
The second slow-roll parameter is defined as
\begin{equation}
    \eta=\dfrac{V''(\phi)}{V(\phi)}M_P^2\,.
\end{equation}
The two most relevant observables characterizing measurements of the CMB are the scalar spectral tilt and the tensor to scalar ratio, which describe the departure from scale invariance at CMB scales and the intensity of tensor perturbations. They can be computed in terms of the slow-roll parameters as
\begin{align}
 n_s-1 &= -6\varepsilon+2\eta\,,\\
 r&=16\varepsilon\,.
\end{align}
For a potential with a general form
\begin{equation}
   V(\phi)=\dfrac{M_P^2M_*^2}{8}e^{2\gamma\phi/M_P}\left(1-e^{-\lambda\phi/M_P}\right)^2
\end{equation}
we can compute without any additional approximation
\begin{align}
     \varepsilon &=2\left(\dfrac{ \gamma  e^{\lambda  \phi/M_P }-\gamma +\lambda}{e^{\lambda  \phi/M_P }-1}\right)^2\,,\\
    \eta &= \frac{4 \gamma ^2 e^{2 \lambda  \phi/M_P }- 2 (-2 \gamma +\lambda )^2 e^{\lambda  \phi/M_P }+4 (-\gamma +\lambda )^2}{\left(e^{\lambda  \phi/M_P }-1\right)^2}\,,\\
    P_\zeta &= \dfrac{\left(e^{\lambda  \phi/M_P }-1\right)^4 e^{2 \phi  (\gamma -\lambda )/M_P}}{384 \pi ^2 \left(\gamma  e^{\lambda  \phi/M_P }-\gamma +\lambda \right)^2}\,.
\end{align}
One can also compute the number of e-folds before the end of inflation in terms of the field,
\begin{equation}
    N_e=\int_{\phi_\text{end}}^{\phi_{N_e}} \dfrac{\dd\phi}{M_P\sqrt{2\varepsilon}}=\dfrac{\lambda  \log \left(\gamma  \left(1-e^{\lambda  \phi/M_P }\right)-\lambda \right)-\gamma  \log \left(\lambda  (\lambda-\gamma ) e^{\lambda  \phi/M_P  }\right)}{2 \gamma  \lambda  (\lambda-\gamma )}\bigg\vert_{\phi_\text{end}}^{\phi_{N_e}}\,.
\end{equation}
In Starobinsky, and in most other models of inflation, one usually neglects the evaluation at $\phi_\text{end}\simeq 0$. Here however, it is relevant in order to appropriately subtract integration constants that arise from the logarithms, which diverge in the limit $\gamma\to 0$. We expect that slow-roll will only be possible for this modified Starobinsky potential exactly when $\gamma$ is sufficiently small. Therefore, we may perform a series expansion in $\gamma\ll 1$ to obtain the first order correction to $\phi_{N_e}$, i.e. the value of $\phi$ at $N_e$ e-folds before the end of inflation.
\begin{equation}
    \phi_{N_e}=\dfrac{\log \left(2 N_e \lambda^2 \left(1 +\gamma N_e \lambda\right)\right)}{\lambda }\,.
\end{equation}
Then we compute the spectral tilt and tensor to scalar ratio at this point. The leading order corrections to the inflationary observables appear at $\mathcal{O}(\gamma)$,
\begin{align}
    n_s-1&= -\frac{2}{N_e} + 2 \gamma \lambda + \mathcal{O}(\gamma^2)\,,\\
    r&= \frac{8}{N_e^2\lambda^2}+\frac{16\gamma}{N_e\lambda} + \mathcal{O}(\gamma^2)\,.
\end{align}
Recall that $\lambda=\sqrt{\frac{2}{3}}$ for the Starobinsky model. The scalar power spectrum takes the form
\begin{equation}
    P_{\zeta}=\dfrac{\lambda ^2 M_*^2 N_e^2}{96 \pi ^2}-\dfrac{\gamma  \lambda ^3 M_*^2 N_e^3}{48 \pi ^2} + \mathcal{O}(\gamma^2)\,.
\end{equation}
Turning this around as a prediction for the value of $M_*$ implied by observations, one finds
\begin{equation}
    M_*^2\simeq \dfrac{96 \pi ^2 P_{\zeta}}{\lambda ^2 N_e^2 (1-2 \gamma  \lambda  N_e)}\,.
\end{equation}
This depends on the choice of $\gamma$, which may rely on which tower becomes light as the inflaton journeys super-Planckian distances. But suppose that it fits within the window $\gamma\in [-0.004,0.001]$ as stated in \eqref{eq:gammarang} to align with observations. Then
\begin{equation}
 0.85   \leq M/M_*\leq 1.05
\end{equation}
where $M_*$ is the expected normalization from Starobinsky. This still is not a big difference (within $\mathcal{O}(1)$), but the scale $M$ can be lowered slightly more than previously expected.

\section{\texorpdfstring{$\Ls$}{Ls} and \texorpdfstring{$R^2$}{R2} from string compactifications}\label{app:strings}
First let us consider the characteristic string- and KK scales when we compactify string theory from $D=4+p$ dimensions to $d=4$ dimensions. The way these are typically deduced is by comparing the Einstein-Hilbert terms
\begin{equation}
\begin{split}
    S &\sim M_s^{2+p}\int \dd^{4+p}x\ \sqrt{-G^{(S)}}\ e^{-2\Phi} G^{(S)MN}\mathcal{R}_{MN}^{(S)} \\
    &\sim \underbrace{\frac{M_s^{2+p}}{g_s^2}\tau^{(E)}_p}_{M_{P,4}^2} \int \dd^4x\ \sqrt{-g^{(E)}}\ g^{(E)\mu\nu}R_{\mu\nu}^{(E)}\,.
\end{split}
\end{equation}
Here $G_{MN}$, $\mathcal{R}_{MN}$ and $g_{\mu\nu}$, $R_{\mu\nu}$ respectively denote the $D$-dimensional and 4-dimensional metric and Ricci tensor. Superscripts refer to the usual string- and Einstein frames, and $\tau^{(E)}_p$ is the Einstein frame volume of an internal $p$-cycle. For example $p=6$ corresponds to compactifying on the entire Calabi-Yau three-fold and $\tau_6^{(E)}=\mathcal{V}^{(E)}$ is the volume modulus. In different literature the volume is often either measured with respect to $M_{P,4}$ (4d Planck units) or $\ell_s=1/M_s$ (string units). Depending on which convention is used, the species scale and species number obey different scaling relations, but inevitably lead to the same results. Here we will stick to string units, where it is useful to define the dimensionless volume $\tau_{p;s}\equiv \tau_p^{(E)}\ell_s^{-p}$. By comparison we recover for instance the usual relations
\begin{equation}\label{eq:stringscale}
    M_s\sim g_{s}^{1/4}M_{P,10} \sim g_s\frac{M_{P,4}}{\sqrt{\tau_{p;s}}}\,.
\end{equation}

\subsection*{Leading KK tower}
For a KK-like tower, the scale at which gravity becomes strongly coupled in the full theory is at the ten-dimensional Planck mass, $M_{P,10}\sim g_s^{-1/4}M_s$. This, together with eq.~\eqref{eq:stringscale} provides the species scale
\begin{equation}
    \Ls \sim M_{P,10}\sim g_s^{3/4}\frac{M_{P,d}}{\sqrt{\tau_{p;s}}}\sim \frac{M_{P,d}}{\sqrt{\tau_{p;s}}}\,.
\end{equation}
One can also rephrase this in terms of the number of species below the cutoff. Let us consider the case where the KK tower is associated to the volume of a certain $p$-cycle becoming large. Allow us furthermore to assume for simplicity an isotropic compactification as explained in the main text. In that case one can write $\mKK\sim R^{-1}\sim \tau_{p;s}^{-1/p}M_s$. The total number of species gives
\begin{equation}
    N\sim \left( \frac{M_{P,10}}{\mKK} \right)^p\sim \tau_{p;s}\,.
\end{equation}
One then recovers the usual convention-independent formula
\begin{equation}
    \Ls\sim \frac{M_{P,4}}{\sqrt{N}}\,.
\end{equation}
In general, starting from $D=d+p$ dimensions and compactifying to $d$ dimensions, one finds
\begin{align}
    \Ls&\sim M_{P,D}\sim \frac{M_{P,d}}{\tau_{p;s}^\frac{1}{d-2}} \sim \dfrac{M_{P,d}}{N^\frac{1}{d-2}}\,,\\[10pt]
    N&\sim \left(\frac{M_{P,D}}{\mKK}\right)^{p}\sim \tau_{p;s}\,.
\end{align}
\subsection*{Leading string tower}
For the case of a leading string tower we will stick to string units. Here, the species scale is identified with the string scale, for which we have
\begin{equation}
    \Ls\sim M_s\sim g_s\dfrac{M_{P,4}}{\sqrt{\mathcal{V}_{s}}}\sim g_s M_{P,4}\,.
\end{equation}
In the last step, we have used the results of \cite{Lee:2019wij}, which argue that leading string towers can only occur at finite volume. This relates the string coupling to the number of species as
\begin{equation}
    N=\dfrac{1}{g_s^2}\,.
\end{equation}
The above is the expected result for towers of excited tensionless strings at weak coupling, which holds also for general $d$ non-compact dimensions
\cite{Dvali:2010vm,Dvali:2012uq}.

\subsection*{Type IIB}
Next we detail how the Starobinsky term $R^2/M^2$ shows up when compactifying the Type IIB string action. Here we take into account leading $\alpha'$ corrections, which appear at cubic order in Type IIB. Starting in string frame the relevant terms are
\begin{equation}
    S_\text{IIB} = \frac{M_s^8}{2}\int \dd^{10}x\ \sqrt{-G^{(S)}}\ e^{-2\Phi}\left[ G^{(S)MN}\mathcal{R}_{MN}^{(S)} + \alpha'^3\mathcal{O}^{MN\cdots}(\mathcal{R}^{(S)4})_{MN\cdots} + \cdots \right]\,.
\end{equation}
Here $\mathcal{O}^{MN\cdots}$ and $(\mathcal{R}^{(S)4})_{MN\cdots}$ are respectively some operators built out of the metric tensor and four Ricci/Riemann curvature tensors. For instance, the Starobinsky term appears from terms like
\begin{equation}
    \mathcal{O}^{MN\cdots}(\mathcal{R}^{(S)4})_{MN\cdots} \supset c_1\left(G^{(S)MN}\mathcal{R}_{MN}^{(S)}\right)^4 \supset c_2 \left( g^{(S)\mu\nu}_4R_{\mu\nu}^{(S)} \right)^2\left( g^{(S)mn}_6R_{mn}^{(S)} \right)^2
\end{equation}
up to some numerical factors, where we have made the index splitting $M=(\mu,m)$. Now let us go to Einstein frame
\begin{equation}
\begin{split}
    G^{(S)MN} &= G^{(E)MN}e^{-\frac{\Phi-\Phi_0}{2}}\,,\\
    \sqrt{-G^{(S)}} &= \sqrt{-G^{(E)}}e^{\frac{5}{2}(\Phi-\Phi_0)}\,,\\
    \mathcal{R}_{MN}^{(S)} &= \mathcal{R}_{MN}^{(E)} + \text{terms involving $\nabla_M\Phi$ and $\nabla^M\nabla_M\Phi$}\,,
\end{split}
\label{eq:einstein_string}
\end{equation}
where we use the convention $\Phi_0\equiv \expval{\Phi}$ such that there is no metric ambiguity on the vacuum \cite{ValeixoBento:2023afn}. Consider then the relevant term from the $\alpha'^3$ correction
\begin{equation}
    S_{\text{IIB},\alpha'^3}\sim M_s^8\int \dd^{10}x \ \sqrt{-G^{(E)}}\ e^{\frac{5}{2}(\Phi-\Phi_0)}e^{-2\Phi} \alpha'^3 e^{-2(\Phi-\Phi_0)}\left( G^{(E)MN}\mathcal{R}_{MN}^{(E)}\right)^4.
\end{equation}
Next, we imagine that the dilaton $\Phi$ and volume $\mathcal{V}^{(E)}$ are stabilised, which means that we can set $\Phi=\Phi_0=\ln g_s$. We can perform the integral over the internal manifold, expecting that the scalar curvature with respect to the internal space metric $g_{mn}^{(E)}$ is of order
\begin{equation}
    R^{(E)}_6\sim \frac{1}{(\mathcal{V}^{(E)})^{1/3}} \sim \frac{M_{P,4}^2}{\mathcal{V}^{1/3}_P}\sim \frac{1}{\mathcal{V}^{1/3}_s\ell_s^2}\,.
\end{equation}
Hence
\begin{equation}
    S\sim \frac{M_s^8}{g_s^2} \mathcal{V}_s\ell_s^6 \int \dd^4x\ \sqrt{-g^{(E)}_4}\left[ R_4+ \alpha'^3 \mathcal{V}^{-2/3}_s \ell_s^{-4}
    R^2_4+\cdots \right].
\end{equation}
Finally, using eqs.~\eqref{eq:stringscale} and that $\alpha'\sim \ell_s^2$, we find
\begin{equation}
    S\sim M_{P,4}^2\int\dd^4\sqrt{-g_4^{(E)}}\left[R_4+ g_s^{-2}\mathcal{V}_s^{1/3} \frac{R^2_4}{M_{P,4}^2}+\cdots\right]\,.
\end{equation}
One can therefore identify up to numerical factors the scale of the $R^2$ term
\begin{equation}
    M\sim g_s\frac{M_{P,4}}{\mathcal{V}^{1/6}_s}\,.
\end{equation}

\subsection*{Heterotic}
Compared the above example of Type IIB, heterotic string theory in $D=10$ dimensions receives corrections already at order $\alpha'$. As such the scale $M$ associated to Starobinsky in $d=4$ after compactyfying will be parametrically different compared to the Type IIB case. The appropriate terms of the action are 
\begin{equation}
    S_\text{het} = \frac{M_s^8}{2}\int \dd^{10}x\ \sqrt{-G^{(S)}}\ e^{-2\Phi}\left[ G^{(S)MN}\mathcal{R}_{MN}^{(S)} + \alpha'\mathcal{O}^{MN\cdots}(\mathcal{R}^{(S)2})_{MN\cdots} + \cdots \right].
\end{equation}
The Starobinsky term enters through the term 
\begin{equation}
    \mathcal{O}^{MN\cdots}(\mathcal{R}^{(S)2})_{MN\cdots} \supset d_1\left(G^{(S)MN}\mathcal{R}_{MN}^{(S)}\right)^2 \supset d_2 \left( g^{(S)\mu\nu}R_{\mu\nu}^{(S)} \right)^2\,,
\end{equation}
again up to some numerical factors. As before we then use eq.~\eqref{eq:einstein_string} to express the action in Einstein frame 
\begin{equation}
    S_{\text{het},\alpha'} \sim M_s^8\int \dd^{10}x \ \sqrt{-G^{(E)}} e^{\frac{5}{2}(\Phi-\Phi_0)}e^{-2\Phi} \alpha' e^{-(\Phi-\Phi_0)}\left( G^{(E)MN}\mathcal{R}_{MN}^{(E)}\right)^2\,.
\end{equation}
Subsequently we assume the volume and dilaton to be appropriately stabilized. We finally find
\begin{equation}
    S\sim \frac{M_s^8}{g_s^2}\mathcal{V}_s\ell_s^6 \int \dd^4 x\ \sqrt{-g^{(E)}_4}\left[ R_4 + \alpha' R_4^2 +\cdots \right]\,.
\end{equation}
The prefactor in front of the integral is the usual $M_{P,4}^2$. Recalling that $\alpha'\sim M_s^{-2}$ we can identify
\begin{equation}
    M\sim M_s\,.
\end{equation}
The scale controlling the $R^2$ term in the four-dimensional EFT is exactly the string scale for heterotic string compactifications.

\vspace{0.4cm}

\bibliographystyle{utphys}
\bibliography{refs}
\end{document}